# Chemical abundances in Terzan 7 from UVES spectra

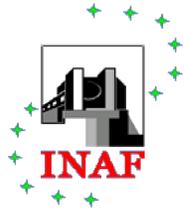
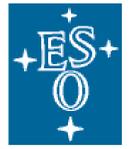

L. Sbordone (ESO – Univ. Roma Tor Vergata), G. Marconi (ESO - OAR),
P. Bonifacio (OAT), R. Buonanno (Univ. Roma Tor Vergata)

## ABSTRACT

We present chemical abundances for Mg, Si, Ca, and Fe for three red giants in the sparse globular cluster Terzan 7, physically associated to the Sagittarius Dwarf Spheroidal Galaxy (Sgr dSph), which is presently being tidally disrupted by the Milky Way. The data has been obtained with VLT-UVES and show a mean [Fe/H]=-0.56, with solar alpha content, mean [α/Fe]=-0.03. This enforces Ter 7's membership to the Sgr dSph system, which shows a similar pattern of abundance.

## INTRODUCTION

The globular cluster Terzan 7 (Buonanno et al. 1995) is one of the 4 clusters strongly associated to the dSph (plus M54, Arp 2, Ter 8) while other six Milky Way clusters have orbital parameters suggesting an origin within the Sgr dSph system, and a consequent stripping. The recent definitive detection of the Sgr tidal stream in the halo, finally, shows the Sgr dSph system as the best available "laboratory" to study the evolution of a dwarf galaxy system in a strongly tidal environment, and its contribution to the halo population.

## OBSERVATIONS AND ANALYSIS

Three red giants have been selected from the RGB of Terzan 7 (see fig 1, lower panel, and physical parameters in table 1) and observed with UVES in dichroic mode 1. Our results come from the analysis of the red arm spectra (central wavelenght 5800 Å, extending from 4800 to 6800 Å), with a S/N ratio of ~30 at central wavelength. The doppler shifted, coadded spectra have been measured using IRAF task splot. The effective temperature have been derived from the calibration of Alonso et al (1999), and the gravity from confrontation with suitable isochrones of the Padova database (Salasnich et al. 2000). Ad hoc model atmosphere have been computed by using ATLAS 9 (Kurucz 1993) ad the abundances for FeI, FeII, Mg, Si, Ca derived from the measured equivalent widths with the WIDTH code (Kurucz 1993) and by direct comparison with synthetic spectra produced by SYNTHE (Kurucz 1993). The iron ionization equilibrium have been used to check the applied gravity.

| Star | v | (V-I)$_0$ | Teff | Log G | |
|---|---|---|---|---|---|
| Ter 7 A | 16.08 | 1.324 | 4168 | 1.30 | 1.50 |
| Ter 7 B | 16.76 | 1.156 | 4436 | 1.80 | 1.45 |
| Ter 7 C | 16,62 | 1.188 | 4385 | 1.60 | 1.40 |

Table 1: physical parameters for the three stars. Colors are dereddened Cousins.

## ABUNDANCES

The derived abundances and uncertainties are listed in tab 2. The stars show high metallicity and no enhancement of alpha-process elements. The observed pattern is coherent with what we found in our analysis of 12 red giants belonging to the Sgr dSph (Bonifacio et al. 2000, Bonifacio et al. 2003, **see the poster at JD 15 by Sbordone et al.**), where we observed high metallicities ([Fe/H]~-0.2) and low alpha abundance compared to Iron, with a tendency towards lower alpha abundance for higher metallicities. In fig. 2 the [alpha/Fe] against [Fe/H] is showed for the three Terzan 7 stars, together with the 12 Sgr dSph stars and 5 M 54 stars studied by Brown et al (1999).

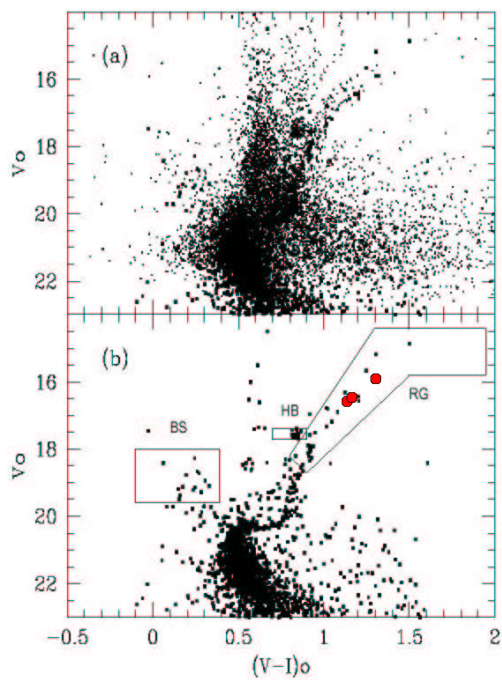

Figure 1: The Terzan 7 CMD (lower panel) and its superposition with the Sgr dSph CMD (Marconi et al 1998). The three stars we observed are marked as red dots.

| Star | [Fe/H] | [Mg/Fe] | [Si/Fe] | [Ca/Fe] |
|---|---|---|---|---|
| Ter 7 A | -0.51 ± 0.11 | 0.08 | 0.02 | -0.27 |
| Ter 7 B | -0.60 ± 0.09 | 0.00 | 0.15 | -0.03 |
| Ter 7 C | -0.61 ± 0.08 | -0.04 | -0.07 | -0.01 |

Table 2: abundance ratios for the three stars.

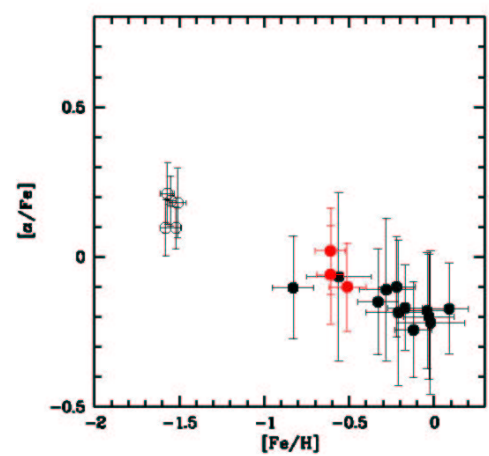

Fig 2: Abundance pattern of Terzan 7 stars (red) together the Sgr dSph sample of Bonifacio et al. (2003) (black filled) and the M54 stars observed by Brown et al (1999) (black open).

M 54 is also associated to the Sgr dSph, of which is suspected to constitute the nucleus. Three more Terzan 7 stars have been observed with UVES by Wallerstein et al. (2002), using a different line set and different $T_{eff}$ scale, showing an excellent agreement with our iron abundances, but a slightly higher alpha content.

## INTERPRETATION

Marconi et al. (1998) noticed strong similarities between the CMD of Terzan 7 and of the Sgr dSph (see fig 1, upper panel). Actually (see Bonifacio et al. 2003, Sbordone et al. 2003) Sgr dSph CMD is biased by strong contamination and heavy age-metallicity degeneracy, due to its prolonged star formation history. Terzan 7, most likely, is constituted by a far simpler population, but seems to lie on the same evolutionary path. This confirms the link with the Sgr dSph (such a low alpha content is unusual for a halo GC). Our findings seem to enforce a scenario in which star formation in the system began more than 10 Gyr ago (M 54 population, for instance) and then continued at slow rate up to the youngest Sgr dSph population. Terzan 7 had apparently formed more or less midway, at least in the sense of the chemical enrichment. The Galaxy tidal shocking is the "prime suspect" for inducing a star formation so much more prolonged with respect to the others local group dwarf galaxies.

## FUTURE PROSPECTS

The scenario here presented for the Sgr system evolution is still highly speculative in many aspects. To enforce or discard it we are planning observations of the (probably) stripped Sgr clusters (NGC 4147, NGC 5634, Pal 12…), together to a more detailed observation of the galaxy itself. As a first step, we plan to re-analyze the three stars observed by Wallerstein et al. (2002) using the same models and procedures we employed for our three giants, to enlarge our sample and verify if the differences in alpha abundances are real or analysis-dependent.